\documentclass[%
reprint,
superscriptaddress,
showpacs,
amsmath,amssymb,
prc,
floatfix, ]%
{revtex4-1}

\usepackage{color}
\usepackage{CJK}

\usepackage{graphicx}
\usepackage{dcolumn}
\usepackage{bm}
\usepackage[dvipdfm,bookmarks=true,colorlinks,%
            citecolor=blue,linkcolor=blue,hypertex, %
            breaklinks=true]{hyperref}

\begin{document}

\begin{CJK*}{GBK}{}


\title{Rotational properties of the superheavy nucleus $^{256}$Rf
       and its neighboring even-even nuclei
       in particle-number conserving cranked shell model}

\author{Zhen-Hua Zhang}%
 \affiliation{State Key Laboratory of Nuclear Physics and Technology,
              School of Physics, Peking University, Beijing 100871, China}
\author{Jie Meng}%
 \affiliation{State Key Laboratory of Nuclear Physics and Technology,
              School of Physics, Peking University, Beijing 100871, China}
 \affiliation{School of Physics and Nuclear Energy Engineering,
              Beihang University, Beijing 100191, China}
 \affiliation{Department of Physics, University of Stellenbosch,
              Stellenbosch 7602, South Africa}
\author{En-Guang Zhao}%
 \affiliation{State Key Laboratory of Theoretical Physics,
              Institute of Theoretical Physics, Chinese Academy of Sciences,
              Beijing 100190, China}
 \affiliation{State Key Laboratory of Nuclear Physics and Technology,
              School of Physics, Peking University, Beijing 100871, China}
 \affiliation{Center of Theoretical Nuclear Physics, National Laboratory
              of Heavy Ion Accelerator, Lanzhou 730000, China}
\author{Shan-Gui Zhou}
 \email{sgzhou@itp.ac.cn}
 \affiliation{State Key Laboratory of Theoretical Physics,
              Institute of Theoretical Physics, Chinese Academy of Sciences,
              Beijing 100190, China}
 \affiliation{Center of Theoretical Nuclear Physics, National Laboratory
              of Heavy Ion Accelerator, Lanzhou 730000, China}

\date{\today}

\begin{abstract}
The ground state band was recently observed in the superheavy nucleus $^{256}$Rf.
We study the rotational properties of $^{256}$Rf and its neighboring
even-even nuclei by using a cranked shell model (CSM)
with the pairing correlations treated by a particle-number conserving (PNC) method
in which the blocking effects are taken into account exactly.
The kinematic and dynamic moments of inertia of the ground state bands
in these nuclei are well reproduced by the theory.
The spin of the lowest observed state in $^{256}$Rf is determined
by comparing the experimental kinematic moments of inertia with
the PNC-CSM calculations and agrees with previous spin assignment.
The effects of the high order deformation $\varepsilon_6$ on the
angular momentum alignments and dynamic moments of inertia in these nuclei
are discussed.
\end{abstract}


\maketitle

\end{CJK*}

\section{\label{Sec:Introduction}Introduction}

In recent years, the in-beam spectroscopy of the nuclei with $Z \approx 100$
has become a hot topic~\cite{Leino2004_ARNPS54-175, Herzberg2004_JPG30-R123,
Herzberg2008_PPNP61-674, Ackermann2011_APPB42-577, Greenlees2011_APPB42-587}.
These transfermium nuclei bring important information for the structure of superheavy nuclei.
Experimental results show that these nuclei are well deformed.
Due to deformation effects, the orbitals originating from
spherical subshells which are important to the magic number in
superheavy nuclei may come close to the Fermi surface in these deformed nuclei.
For example, the $\pi 1/2^-[521]$ and $\pi 3/2^-[521]$ orbitals are of
particular interest since they stem from the spherical
$\pi2f_{5/2,7/2}$ orbtials; the splitting between these spin doublets is very
important to the location of the next proton shell closure.
The high spin rotational states of these transfermium nuclei can give valuable
information about the single particle orbitals near the Fermi surface,
especially the high-$j$ intruder orbitals ($\nu j_{15/2}$ or $\pi i_{13/2}$)
which are sensitive to the Coriolis interaction.

In a recent work~\cite{Zhang2012_PRC85-014324},
the spectroscopy of the nuclei with $Z\approx 100$ is systematically
investigated by a particle-number conserving (PNC) method
based on a cranked shell model (CSM)~\cite{Zeng1983_NPA405-1, Zeng1994_PRC50-1388}
with a new Nilsson parameter set which is obtained by fitting the
experimental single-particle spectra in these nuclei.
The calculated bandhead energies of the one-quasiproton and one-quasineutron
bands in odd-$A$ nuclei are improved dramatically comparing with those calculated by
using the traditional Nilsson parameter~\cite{Nilsson1969_NPA131-1}.
In contrary to the conventional Bardeen-Cooper-Schrieffer (BCS) or
Hartree-Fock-Bogolyubov (HFB) approach, in the PNC method, the CSM Hamiltonian
is solved directly in a truncated Fock-space~\cite{Wu1989_PRC39-666}.
Therefore the particle-number is conserved and the Pauli blocking effects
are taken into account exactly.
The experimental kinematic moments of inertia (MOI's) for the rotational
bands in even-even, odd-$A$, and odd-odd nuclei with $Z \approx 100$ are reproduced
quite well by the PNC-CSM calculations.
The PNC scheme has also been implemented both in relativistic
and nonrelativistic mean field models~\cite{Meng2006_FPC1-38, Pillet2002_NPA697-141}
in which the single-particle states are calculated from self-consistent
mean field potentials instead of the Nilsson potential.

Quite recently, the ground state bands (GSB)
were observed in the even-even nuclei $^{246}$Fm ($Z=100$)~\cite{Piot2012_PRC85-041301R}
and $^{256}$Rf ($Z=104$)~\cite{Greenlees2012_PRL109-012501}.
It is worthwhile to mention that Rf is the first element whose
stability is entirely due to the quantum shell effects
and it marks the gateway to superheavy elements~\cite{Schadel2006_AngewChem_Int45-368}.
The spectrum and MOI's of $^{256}$Rf can give information about
the single-particle structure and the pairing interaction of the
superheavy nuclei and provide a test for current nuclear models.
$^{246}$Fm has been included in our systematic investigation~\cite{Zhang2012_PRC85-014324, Zhang2012_PhD}.
In this paper, we extend the PNC-CSM to the study of the rotational
properties of $^{256}$Rf and its neighboring even-even nuclei.
The spin of the experimentally observed lowest-lying state in
$^{256}$Rf will be determined by comparing the kinematic MOI's
with the PNC-CSM calculations.
We further study the effects of high order deformation $\varepsilon_6$
on the angular momentum alignment and dynamic MOI's in these nuclei.

The paper is organized as follows.
A brief introduction of the PNC-CSM is presented in Sec.~\ref{Sec:PNC-CSM}.
The results and discussions are given in Sec.~\ref{Sec:Results}.
Finally we summarize our work in Sec.~\ref{Sec:Summary}.

\section{\label{Sec:PNC-CSM}Theoretical framework}

The cranked Nilsson Hamiltonian of an axially symmetric nucleus in the rotating
frame can be written as
\begin{eqnarray}
 H_\mathrm{CSM}
 & = &
 H_0 + H_\mathrm{P}
 = H_{\rm Nil}-\omega J_x + H_\mathrm{P}
 \ ,
 \label{eq:H_CSM}
\end{eqnarray}
where $H_{\rm Nil}$ is the Nilsson Hamiltonian, $-\omega J_x$
is the Coriolis interaction with the cranking frequency
$\omega$ about the $x$ axis (perpendicular to the nuclear
symmetry $z$ axis). $H_{\rm P} = H_{\rm P}(0) + H_{\rm P}(2)$
is the pairing interaction,
\begin{eqnarray}
 H_{\rm P}(0)
 & = &
  -G_{0} \sum_{\xi\eta} a^\dag_{\xi} a^\dag_{\bar{\xi}}
                        a_{\bar{\eta}} a_{\eta}
  \ ,
 \\
 H_{\rm P}(2)
 & = &
  -G_{2} \sum_{\xi\eta} q_{2}(\xi)q_{2}(\eta)
                        a^\dag_{\xi} a^\dag_{\bar{\xi}}
                        a_{\bar{\eta}} a_{\eta}
  \ ,
\end{eqnarray}
where $\bar{\xi}$ ($\bar{\eta}$) labels the time-reversed state of a
Nilsson state $\xi$ ($\eta$), $q_{2}(\xi) = \sqrt{{16\pi}/{5}}
\langle \xi |r^{2}Y_{20} | \xi \rangle$ is the diagonal element of
the stretched quadrupole operator, and $G_0$ and $G_2$ are the
effective strengths of monopole and quadrupole pairing interactions,
respectively.

Instead of the usual single-particle level truncation in conventional
shell-model calculations, a cranked many-particle configuration
(CMPC) truncation (Fock space truncation) is adopted~\cite{Zeng1994_PRC50-1388,
Molique1997_PRC56-1795}.
An eigenstate of $H_\mathrm{CSM}$ can be written as
\begin{equation}
 |\Psi\rangle = \sum_{i} C_i \left| i \right\rangle
 \ ,
 \qquad (C_i \; \textrm{real}),
\end{equation}
where $| i \rangle$ is a CMPC (an eigenstate of the one-body operator $H_0$).
By diagonalizing the $H_\mathrm{CSM}$ in a sufficiently
large CMPC space, sufficiently accurate solutions for low-lying excited
eigenstates of $H_\mathrm{CSM}$ are obtained~\cite{Zhang2012_PRC85-014324}.
The angular momentum alignment for the state $| \Psi \rangle$ is
\begin{equation}
 \left\langle \Psi | J_x | \Psi \right\rangle
 =\sum_i C_i^2 \left\langle i | J_x | i \right\rangle
 +2\sum_{i<j} C_i C_j \left\langle i | J_x | j \right\rangle \ .
\end{equation}
Considering $J_x$ to be a one-body operator, the matrix element
$\langle i | J_x | j \rangle$ for $i\neq j$ is nonzero only when
$|i\rangle$ and $|j\rangle$ differ by one particle
occupation~\cite{Zeng1994_PRC50-1388}.
After a certain permutation of
creation operators, $|i\rangle$ and $|j\rangle$ can be recast into
\begin{equation}
 | i \rangle = (-1)^{M_{i\mu}} | \mu \cdots \rangle \ , \qquad
 | j \rangle = (-1)^{M_{j\nu}} | \nu \cdots \rangle \ ,
\end{equation}
where the ellipsis ``$\cdots$''
stands for the same particle occupation and
$(-1)^{M_{i\mu(\nu)}}=\pm1$ according to whether the permutation is
even or odd. Therefore, the kinematic MOI  $J^{(1)}$ of
$|\Psi\rangle$ can be separated into the diagonal and the
off-diagonal parts
\begin{eqnarray}
 J^{(1)}
  &=& \frac{1}{\omega} \left\langle \Psi | J_x | \Psi \right\rangle
  = \frac{1}{\omega}
  \left( \sum_{\mu} j_x(\mu) + 2\sum_{\mu<\nu} j_x(\mu\nu) \right) \ ,\\
 j_x(\mu)
 &=& \langle \mu | j_{x} | \mu \rangle n_{\mu}  \ , \label{eq:j1d} \\
 j_x(\mu\nu)
 &=&\langle \mu | j_{x} | \nu \rangle
  \sum_{i<j} (-1)^{M_{i\mu}+M_{j\nu}} C_{i} C_{j} \ ,
  \label{eq:j1od}
\end{eqnarray}
where $n_{\mu} = \sum_{i} |C_{i}|^{2} P_{i\mu}$ is the
occupation probability of the cranked Nilsson orbital $|\mu\rangle$
and $P_{i\mu}=1$ (0) if $|\mu\rangle$ is occupied (empty).
The expression of the dynamic MOI
$J^{(2)} = d \left\langle \Psi | J_x | \Psi \right\rangle / d \omega $
is similar.

\section{\label{Sec:Results}Results and discussions}

The parameters used in this work are all taken from
Ref.~\cite{Zhang2012_PRC85-014324}.
We note that due to the velocity-dependent $l^2$ term,
the MOI's of very high-spin states can not be well
described by a cranked Nilsson model~\cite{
Brack1976_NPA258-264, Neergard1976_NPA262-61,
Andersson1976_NPA268-205,
Neergard1977_NPA287-48,
Bengtsson1978_PS17-583,
Bengtsson1985_NPA436-14,
Bengtsson1987_NPA473-77,
Bengtsson1989_PS39-196}.
However, we are mainly focusing on relatively low-spin states.
The MOI's are mainly determined by the pairing interaction
and the single particle levels near the Fermi surface,
especially the location of the high-$j$ intruder orbitals.
As will be seen in the following, the calculated MOI's
with our model agree well the experiment.
The traditional Nilsson parameters ($\kappa$ and $\mu$)
~\cite{Nilsson1969_NPA131-1, Bengtsson1985_NPA436-14}
are optimized to reproduce the experimental level schemes
for the rare-earth and actinide nuclei near the stability line.
However, this parameter set can not describe well the
experimental level schemes of transfermium nuclei.
Therefore the new set of Nilsson parameters ($\kappa$ and $\mu$)
obtained by fitting the experimental single-particle spectra
in these nuclei in Ref.~\cite{Zhang2012_PRC85-014324} is adopted here.
Note that this set of parameters has been used to study
rotational bands in $^{247,249}${Cm} and $^{249}${Cf} in Ref.~\cite{Zhang2011_PRC83-011304R}.

The experimental values of the deformation parameters for the transfermium
nuclei are very scare and the predictions of different theories are not
consistent with each other~\cite{Moeller1995_ADNDT59-185,
Sobiczewski2001_PRC63-034306, Afanasjev2003_PRC67-024309}.
In the PNC-CSM calculations, the deformations are chosen to be close
to existing experimental values and change smoothly with
the proton and the neutron numbers.
The deformation parameters of $^{256}$Rf can be extrapolated
from Table~II in Ref.~\cite{Zhang2012_PRC85-014324} as
$\varepsilon_2=0.255$ and $\varepsilon_4=0.025$.

The CMPC space in this work is constructed in the proton $N=4, 5, 6$ shells
and the neutron $N=6, 7$ shells.
The dimensions of the CMPC space are about 1000 both for protons and neutrons.
The effective pairing strengths are $G_p$ = 0.40~MeV, $G_{2p}$ = 0.035~MeV,
$G_n$ = 0.30~MeV, and $G_{2n}$ = 0.020~MeV, which are the same for all even-even
nuclei in this mass region (see Table~III in Ref.~\cite{Zhang2012_PRC85-014324}).

\begin{figure}
\includegraphics[width=0.99\columnwidth]{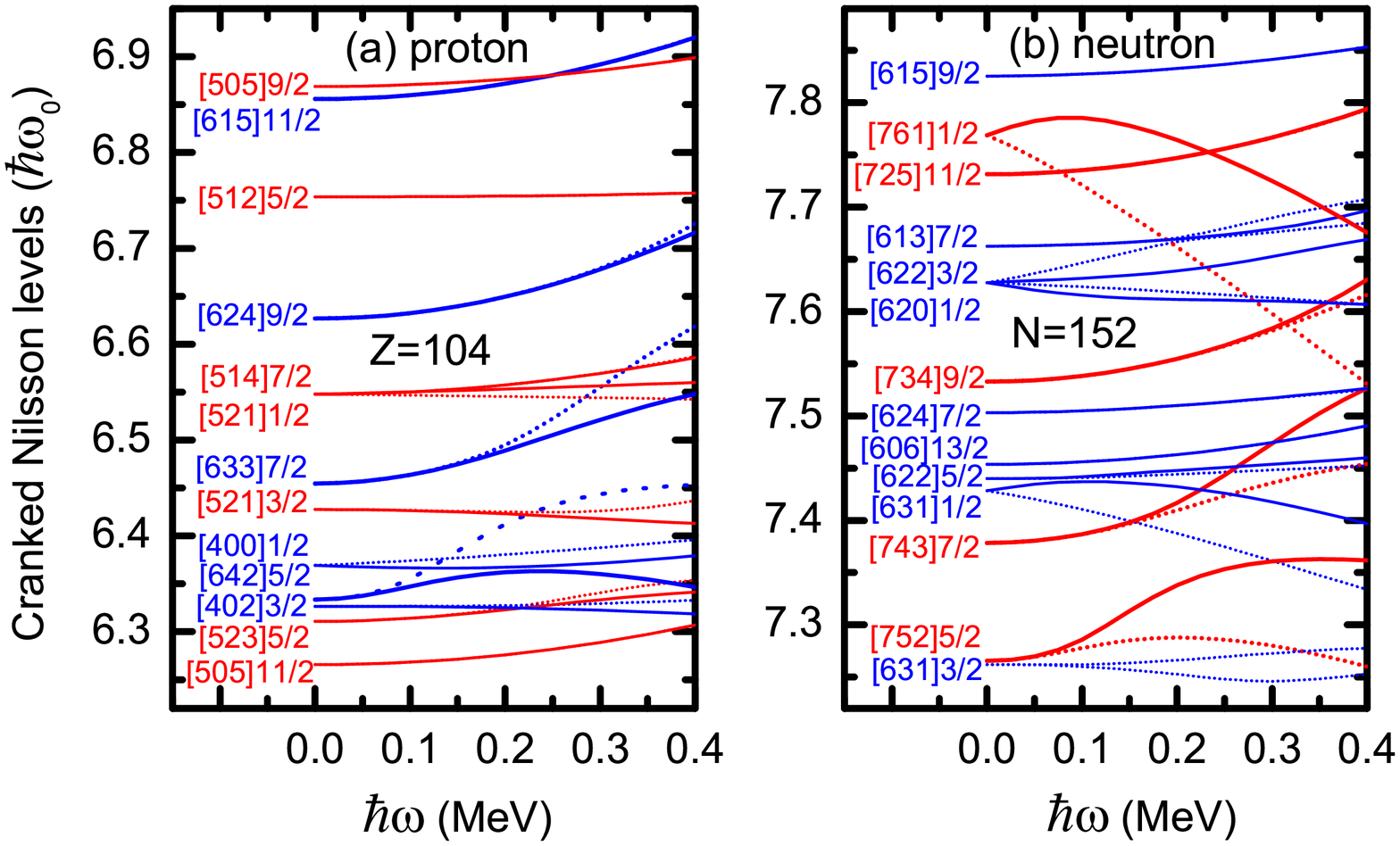}
\caption{\label{fig:Nilsson}
(Color online)
The cranked Nilsson levels near the Fermi surface of $^{256}$Rf
(a) for protons and (b) for neutrons. The positive (negative) parity
levels are denoted by blue (red) lines. The signature $\alpha=+1/2$
($\alpha=-1/2$) levels are denoted by solid (dotted) lines.
The deformation parameters $\varepsilon_2= 0.255$ and $\varepsilon_4=0.025$.
}
\end{figure}

Figure~\ref{fig:Nilsson} shows the calculated cranked Nilsson levels
near the Fermi surface of $^{256}$Rf. The positive (negative)
parity levels are denoted by blue (red) lines.
The signature $\alpha=+1/2$ ($\alpha=-1/2$) levels are
denoted by solid (dotted) lines.
From Fig.~\ref{fig:Nilsson} it can be seen that there exist a proton gap at
$Z=100$ and a neutron gap at $N=152$, which is consistent with the
calculation by using Woods-Saxon potential~\cite{Chasman1977_RMP49-833,
*Chasman1978_RMP50-173, Sobiczewski2001_PRC63-034306}.
The $Z=104$ proton energy spacing in our calculation is about 0.5~MeV,
which is much larger than that of $Z=102$.
This situation is the opposite in Ref.~\cite{Sobiczewski2001_PRC63-034306}.
Nevertheless, the $Z=104$ gap in our calculation is not very significant
comparing with that calculated by the self-consistent mean field
models~\cite{Bender2003_NPA723-354, Afanasjev2003_PRC67-024309},
which is usually larger than 1~MeV.
For protons, the sequence of single-particle levels near the Fermi surface
in our calculation is quite similar with that determined from the
experimental information of $^{255}$Lr~\cite{Chatillon2006_EPJA30-397},
in which the energies of $\pi 1/2^-[521]$ and $\pi 7/2^-[514]$ are nearly degenerate.
For neutrons, it is shown experimentally that the ground state of
$^{255}$Rf is $\nu 9/2^-[734]$~\cite{Hessberger2006_EPJA30-561},
which is also consistent with our calculation.

\begin{figure}[h]
\includegraphics[width=0.90\columnwidth]{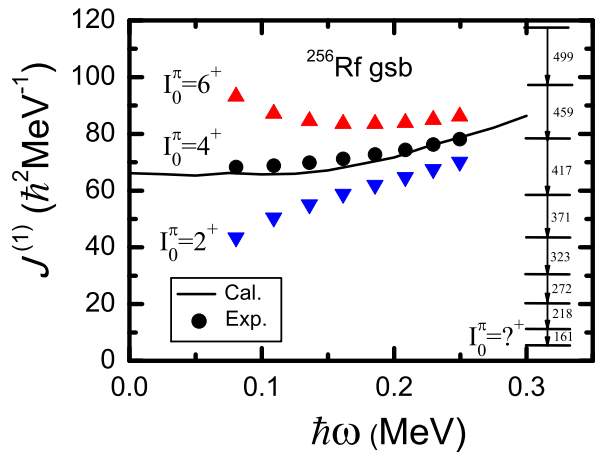}
\caption{\label{fig:256Rf} (Color online)
The comparison of experimental kinematic MOI's of the
GSB in $^{256}$Rf with the PNC-CSM calculations.
The red up-triangles, black solid circles, and blue down-triangles
denote the experimental MOI's extracted by assigning the 161~keV
transition as $8^+\rightarrow 6^+$, $6^+\rightarrow 4^+$,
and $4^+\rightarrow 2^+$, respectively.
The data are taken from Ref.~\cite{Greenlees2012_PRL109-012501}.}
\end{figure}

Because of the dominance of internal conversion,
the lowest $\gamma$ transitions in $^{256}$Rf were not detected and
spins of states in the observed rotational band were not determined experimentally.
There are many ways to make spin assignments by fitting the rotational band
with various empirical rotational formulae or models~\cite{Harris1964_PRL13-663,
*Harris1965_PR0138-509B, Holmberg1968_NPA117-552,
Wu1987_CTP8-51, Zeng1991_PRC44-1745R, Wu1992_PRC45-261, Xu1994_PRC49-1449,
*Xu1995_PRC52-431, Zhou1997_NPA615-229, Zhou1997_PRC55-2324}.
The $ab$-formula~\cite{Holmberg1968_NPA117-552,
Wu1987_CTP8-51, Zeng1991_PRC44-1745R, Wu1992_PRC45-261} and
the Harris formula~\cite{Harris1964_PRL13-663, Harris1965_PR0138-509B}
have been used to assign the spin and to extrapolate the energies corresponding
to unobserved transitions in $^{246}$Fm and $^{256}$Rf.
The spin assignment for the rotational band observed in
$^{253}$No~\cite{Herzberg2009_EPJA42-333} has already been made in
Ref.~\cite{Wen2012_SciSinPMA42-22} by using the $ab$ formula
which supports the configuration assignment of $\nu 7/2^+[624]$
for this rotational band.
The kinematic MOI's depend sensitively on the spin assignment;
this feature can also be used to make spin assignments for those rotational bands
whose spins are not experimentally determined.
In Fig.~\ref{fig:256Rf} we show the comparison of experimental
kinematic MOI's of the GSB in $^{256}$Rf extracted from different
spin assignments with the PNC-CSM calculations.
The red up-triangles, black solid circles, and blue down-triangles
denote the experimental MOI's extracted by assigning
the observed lowest-lying 161~keV transition as $8^+\rightarrow 6^+$,
$6^+\rightarrow 4^+$, and $4^+\rightarrow 2^+$, respectively.
Our calculation agrees very well with the $6^+\rightarrow 4^+$ assignment,
and is also consistent with the spin assignment using the Harris
formula~\cite{Greenlees2012_PRL109-012501}.
So in the following calculations, the 161~keV transition is assigned as
$6^+\rightarrow 4^+$ and the deduced energies of $4^+\rightarrow 2^+$ (104~keV)
and $2^+\rightarrow 0^+$ (44~keV) in Ref.~\cite{Greenlees2012_PRL109-012501}
are also used to calculate the experimental kinematic and dynamic
MOI's in the GSB of $^{256}$Rf.
This method has also been used to make the spin assignment
for the ground state band established in $^{246}$Fm
and the spin of the lowest state (fed by the 167 keV transition)
is determined to be $4\hbar$~\cite{Zhang2012_PhD},
which is consistent with Ref.~\cite{Piot2012_PRC85-041301R}.

\begin{figure}[h]
\includegraphics[width=0.90\columnwidth]{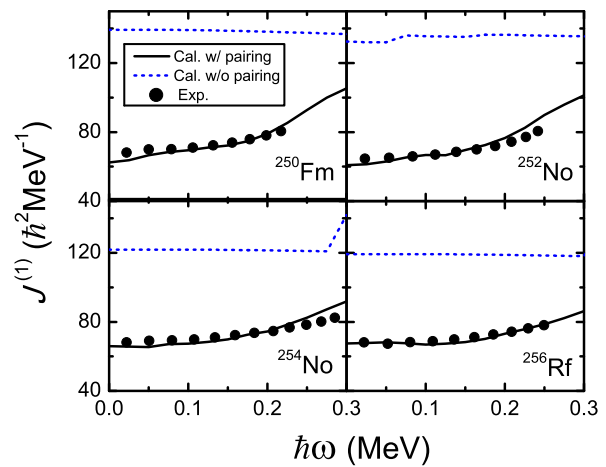}
\caption{\label{fig:J1FmNoRf} (Color online)
The experimental (solid circles) and calculated kinematic MOI's
$J^{(1)}$ with (solid black lines) and without (dotted blue lines)
pairing correlations for $^{256}$Rf and the neighboring even-even nuclei
$^{250}$Fm~\cite{Bastin2006_PRC73-024308} and
$^{252,254}$No~\cite{Herzberg2001_PRC65-014303, Eeckhaudt2005_EPJA26-227}.}
\end{figure}

To study the influence of pairing correlations on rotational properties,
experimental (solid circles) and calculated kinematic MOI's
$J^{(1)}$ with (solid black lines) and without (dotted blue lines)
pairing correlations for $^{256}$Rf and the neighboring even-even nuclei
$^{250}$Fm~\cite{Bastin2006_PRC73-024308} and
$^{252,254}$No~\cite{Herzberg2001_PRC65-014303, Eeckhaudt2005_EPJA26-227}
are shown in Fig.~\ref{fig:J1FmNoRf}.
The pairing interaction is very important in reproducing
the experimental MOI's.  It can be seen that the MOI's of these four
nuclei are roughly overestimated by a factor of two at the bandhead
when the pairing interaction is switched off.
When the pairing interaction is switched on, the observed
MOI's are reproduced quite well, especially for $^{256}$Rf.
This indicates that
the single-particle levels we adopted here are reasonable in this mass region,
which shows that there exist
a proton gap at $Z=100$ and a neutron gap at $N=152$ and
the proton gap at $Z=104$ is not so pronounced.

\begin{figure}[h]
\includegraphics[width=0.90\columnwidth]{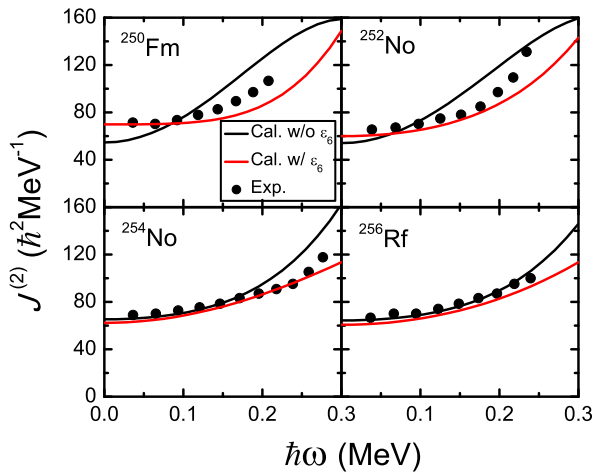}
\caption{\label{fig:J2FmNoRf} (Color online)
The experimental (solid circles) and calculated (solid black lines)
dynamic MOI's $J^{(2)}$ for $^{256}$Rf and the neighboring even-even
nuclei $^{250}$Fm~\cite{Bastin2006_PRC73-024308} and
$^{252,254}$No~\cite{Herzberg2001_PRC65-014303, Eeckhaudt2005_EPJA26-227}.
The red lines are the results when the high order deformation
parameter $\varepsilon_6$ is considered in the PNC-CSM calculation.
$\varepsilon_6$ for $^{250}$Fm, $^{252,254}$No, and $^{256}$Rf
are 0.044, 0.040, 0.042, and 0.038, respectively,
which are taken from Ref.~\cite{Moeller1995_ADNDT59-185}.}
\end{figure}

In Fig.~\ref{fig:J2FmNoRf} we show the experimental (solid circles)
and calculated (solid black lines) dynamic MOI's $J^{(2)}$ for $^{256}$Rf
and the neighboring even-even nuclei $^{250}$Fm~\cite{Bastin2006_PRC73-024308}
and $^{252,254}$No~\cite{Herzberg2001_PRC65-014303,
Eeckhaudt2005_EPJA26-227}.
The experimental dynamic MOI's $J^{(2)}$ for $^{256}$Rf are
reproduced perfectly by the PNC-CSM calculation.
For the other three nuclei the results are
also satisfactory compared with the experiment
though there are some deviations.
As pointed out in Ref.~\cite{Greenlees2012_PRL109-012501},
the alignment of $N=150$ isotones ($^{250}$Fm and $^{252}$No) occurs a
little earlier than that of $N=152$ isotones ($^{254}$No and $^{256}$Rf)
and is delayed in $^{254}$No relative to $^{256}$Rf.
The upbending mechanism in this mass region has been investigated in
detail in our previous work~\cite{Zhang2012_PRC85-014324}.
Similar results have been achieved by other models~\cite{Afanasjev2003_PRC67-024309,
Al-Khudair2009_PRC79-034320}.
However, we can not reproduce the alignment delay in $^{254}$No.
In $N=150$ isotones, more shape degrees of freedom other than
$\varepsilon_{2}$ and $\varepsilon_{4}$, e.g., the $Y_{32}$ correlation
and nonaxial octupole deformation
may play important roles~\cite{Chen2008_PRC77-061305R, Zhao2012_PRC86-057304}.
In particular, Liu et al. explained the fast alignment in $^{252}$No
and slow alignment in $^{254}$No in terms of $\beta_6$ deformation
which decreases the energies of the neutron $j_{15/2}$ intruder
orbitals below the $N = 152$ gap~\cite{Liu2012_PRC86-011301R}.
Here in Fig.~\ref{fig:J2FmNoRf} we show our results for the dynamic MOI's
after considering this high order deformation.
The red lines are the results when $\varepsilon_6$
is considered in the PNC-CSM calculation.
The values of $\varepsilon_6$ for $^{250}$Fm, $^{252,254}$No, and $^{256}$Rf
are 0.044, 0.040, 0.042, and 0.038, respectively,
which are taken from Ref.~\cite{Moeller1995_ADNDT59-185}.
It can be seen that the $\varepsilon_6$ deformation have
prominent effect in the high rotational frequency region.
The results are improved after considering this deformation in the PNC-CSM.
Note that the deformation parameter $\varepsilon_6$ is
fixed in our calculation while
it changes with the rotational frequency in the Total
Routhian Surface (TRS) calculation in Ref.~\cite{Liu2012_PRC86-011301R}.
We expect that after considering this effect,
the results can be improved further.

\begin{figure}[h]
\includegraphics[width=0.90\columnwidth]{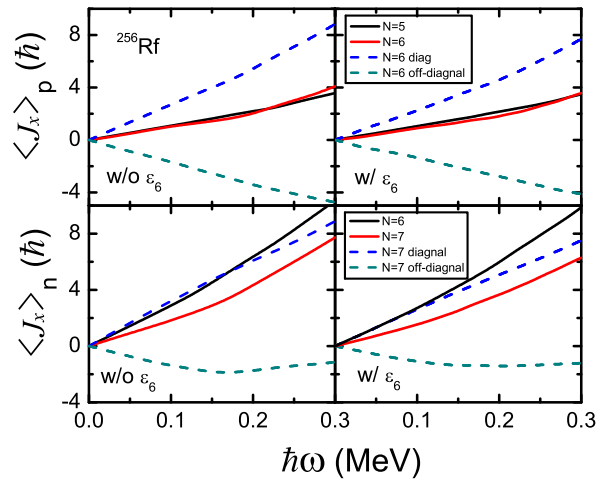}
\caption{\label{fig:Rfjxshell} (Color online)
Contribution of each proton and neutron major shell to
the angular momentum alignment $\langle \Psi | J_x | \Psi \rangle$
for the GSB in $^{256}$Rf.
The left (right) part is the result without (with)
$\varepsilon_6$ deformation.
The diagonal $\sum_{\mu} j_x(\mu)$ calculated from Eq.~(\protect\ref{eq:j1d}) and
off-diagonal parts $\sum_{\mu<\nu} j_x(\mu\nu)$ from
Eq.~(\protect\ref{eq:j1od}) for the proton $N=6$ and neutron $N=7$
shells are shown by dashed lines.%
}
\end{figure}

It should be stressed that although the similar effects of
$\varepsilon_6$ deformation on MOI's are obtained by both TRS
method and PNC-CSM, the upbending mechanisms are different.
The contribution of each proton and neutron major shell to the
angular momentum alignment $\langle \Psi | J_x | \Psi \rangle$ for the GSB in
$^{256}$Rf is shown in Fig.~\ref{fig:Rfjxshell} to illustrate this point.
The left (right) part of Fig.~\ref{fig:Rfjxshell} is the result without (with)
$\varepsilon_6$ deformation.
The diagonal parts $\sum_{\mu} j_x(\mu)$ calculated from Eq.~(\protect\ref{eq:j1d}) and
off-diagonal parts $\sum_{\mu<\nu} j_x(\mu\nu)$ from Eq.~(\protect\ref{eq:j1od})
for the proton $N=6$ and the neutron $N=7$ shells are shown by dashed lines.
It can be seen from the left part of Fig.~\ref{fig:Rfjxshell} that
the alignments of protons and neutrons take
place simultaneously in $^{256}$Rf and the neutron contribution seems
just a little larger than the proton,
which is due to the off-diagonal part of the neutron $N=7$ major shell.
After considering the $\varepsilon_6$ degree of freedom,
both the contribution from protons and neutrons are reduced,
but the competition of alignments still exists.
The conclusion by the TRS method in Ref.~\cite{Liu2012_PRC86-011301R}
is different, which indicate that the neutron $\nu j_{15/2}$ orbital
contributes a lot to the alignment, while the
contribution from proton $\pi i_{13/2}$ is very small.

\begin{figure}[h]
\includegraphics[width=0.95\columnwidth]{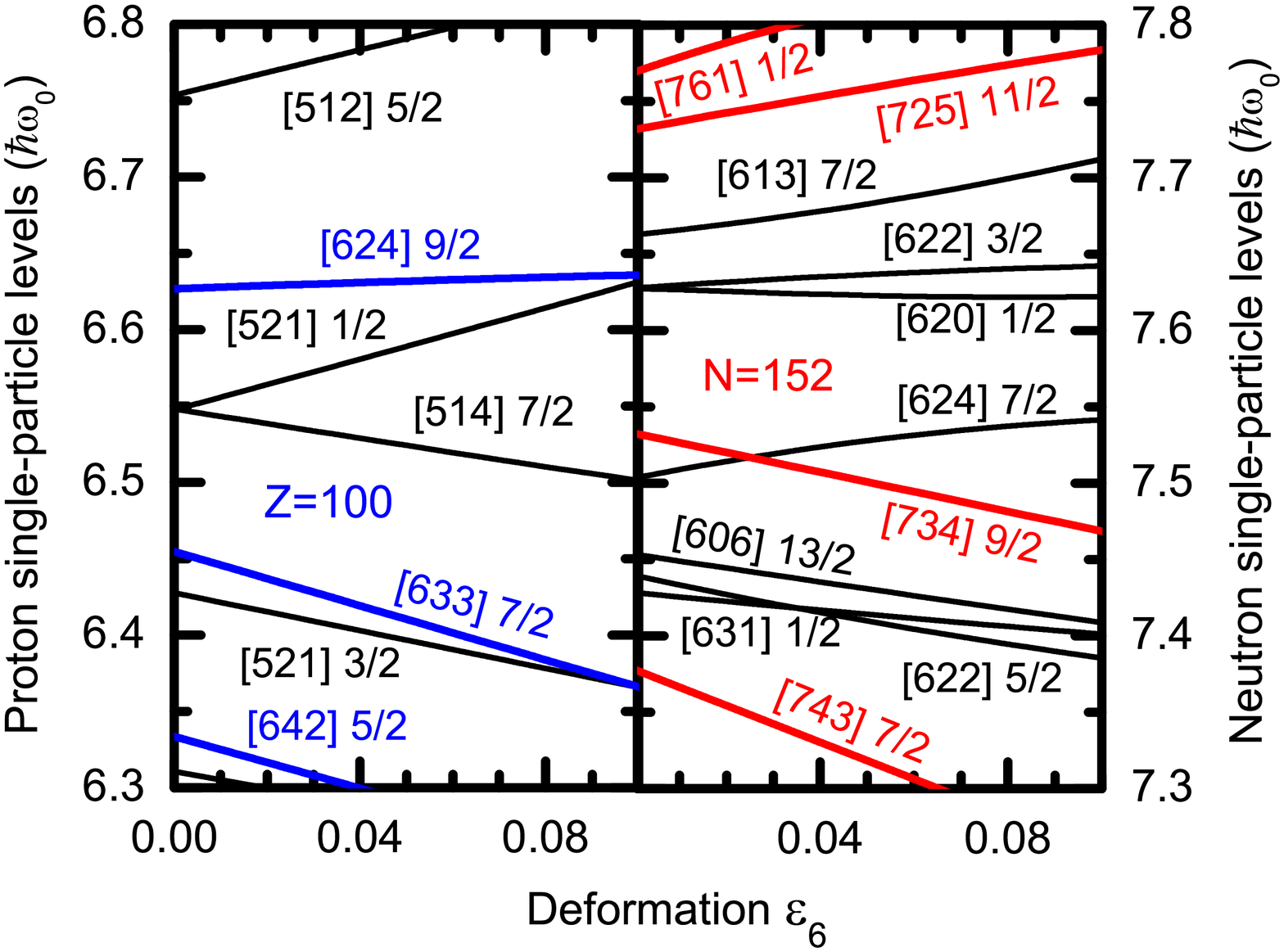}
\caption{\label{fig:RfSPL} (Color online)
The single particle levels near the Fermi surface of $^{256}$Rf
as a function of $\varepsilon_6$ deformation.
The deformation parameters $\varepsilon_2=0.255$ and $\varepsilon_4=0.025$.
The proton and neutron intruder orbital are denoted by
blue and red lines, respectively.
}
\end{figure}

The single particle levels near the Fermi surface of $^{256}$Rf
as a function of $\varepsilon_6$ deformation is shown in Fig.~\ref{fig:RfSPL}.
The proton and neutron intruder orbital are denoted by
blue and red lines, respectively.
It can be seen that the $\varepsilon_6$ deformation lowers not only
the neutron $\nu j_{15/2}$ intruder orbitals below the $N=152$ subshell,
but also the proton $\pi i_{13/2}$ intruder orbitals below the
$Z=100$ subshell. This is the reason why both proton and neutron
contributions to the upbending are reduced when the $\varepsilon_6$
deformation is included in the PNC-CSM.

\section{\label{Sec:Summary}Summary}

The recently observed high-spin rotational ground state band in
$^{256}$Rf~\cite{Greenlees2012_PRL109-012501}
and those in its neighboring even-even nuclei are
investigated by using a cranked shell model with pairing correlations
treated by a particle-number conserving method
in which the blocking effects are taken into account exactly.
Both the experimental kinematic and dynamic MOI's are reproduced
quite well by the PNC-CSM calculations.
The spin of the experimentally observed lowest-lying state in
the GSB of $^{256}$Rf is determined by comparing the MOI's extracted
from different spin assignments with the calculations.
Thus determined spin for the observed lowest-lying state is $4\hbar$ and
consistent with the spin assignment made by using the Harris formula
in Ref.~\cite{Greenlees2012_PRL109-012501}.
We paid much attention to the different rotational behaviors among $^{256}$Rf
and its neighboring even-even nuclei and
the effects of the high order deformation $\varepsilon_6$ on the
angular momentum alignment.
The $\varepsilon_6$ deformation has a noticeable effect on
the dynamical moments of inertia in the high rotational frequency region.
The calculation results for the dynamical moments of inertia are improved
after considering this high order deformation in the PNC-CSM.
In present calculation, the Nilsson potential has been used,
it will be interesting to perform similar investigation with
a Woods-Saxon potential to generate the basis states in the future.

\begin{acknowledgements}

We thank P. T. Greenlees for sending us
the data of $^{256}$Rf prior to publication.
Helpful discussions with P. T. Greenlees, J. Piot, and P. W. Zhao
are gratefully acknowledged.
This work has been supported by
National Key Basic Research Program of China (Grant No. 2013CB834400),
National Natural Science Foundation of China
(Grant No. 10975008, No. 10975100, No. 10979066,
No. 11121403, No. 11175002, No. 11175252, No. 11120101005, No. 11120101005, and No. 11275248),
Knowledge Innovation Project of Chinese Academy of Sciences
(Grant No. KJCX2-EW-N01 and No. KJCX2-YW-N32),
the Research Fund for the Doctoral Program of Higher Education
under Grant No. 20110001110087.
The results described in this paper are obtained on the
ScGrid of Supercomputing Center,
Computer Network Information Center of Chinese Academy of Sciences.

\end{acknowledgements}


%

\end{document}